\documentclass[showpacs,preprintnumbers,amsmath,amssymb,aps,pre]{revtex4}

\usepackage{graphicx,psfrag}
\usepackage{amssymb}
\usepackage{amsmath}
\usepackage{bm}
\usepackage[dvips]{epsfig}
 \graphicspath{{./Figures/}}

\begin{document}

\vspace{1cm}
\title{Rheological properties of  the soft disk model of two-dimensional foams}

\author{Vincent J. Langlois}
\email{vincent.langlois@fysik.dtu.dk}
\affiliation{School of Physics, Trinity College Dublin. Dublin 2.
Ireland. \\
Department of Physics, Technical University of Denmark. 2800 Kgs. Lyngby.
Denmark.}
\author{Stefan Hutzler}
\affiliation{School of Physics, Trinity College Dublin. Dublin 2.
Ireland.}
\author{Denis Weaire}
\affiliation{School of Physics, Trinity College Dublin. Dublin 2.
Ireland.}

\date{\today}


\begin{abstract}

The soft disk model previously developed and applied by Durian [\textit{Phys.
Rev. Lett.}, \textbf{75}:4780-4783 (1995)] is brought to bear on problems of
foam rheology of longstanding and current interest, using two-dimensional (2D)
systems. The questions
at issue include the origin of the Herschel-Bulkley relation, normal stress
effects (dilatancy),
and localisation in the presence of wall drag. We show that even a model
that incorporates only linear
viscous effects at the local level gives rise to nonlinear (power-law) dependence of the limit stress
on strain rate. With wall drag, shear localisation is found.
Its non-exponential form and the variation of localisation length
with boundary velocity are well
described by a continuum model in the spirit of Janiaud {\em et al.}
[\textit{Phys. Rev. Lett.}, \textbf{93}:18303 (2006)].
Other results satisfactorily link localisation to model
parameters, and hence  tie together continuum and local  descriptions, for
the first time. 
\end{abstract}

\maketitle

\section{Introduction}
\subsection{Foam rheology}
While the deformation and flow properties of foams are broadly understood in terms of  shear
elastic modulus, yield stress, etc. \cite{Kraynik88,WeaireFortes94,Hoehler05}, many details remain
perplexing. A fuller understanding  must address both the local forces that operate at the level of
the individual films and the way in which these forces combine to determine the overall response to
strain. This paper will be entirely devoted to the second question. It uses
a particularly simple
representation of bubbles and their mutual forces, as previously developed by Durian
\cite{Durian95,Durian97,Tewari99,Ono03}. 
This model may be unrealistic in some respects, but its simplicity and computational tractability
makes it attractive at a time when more precise descriptions of dynamic properties are lacking.

Our
immediate goal is to thoroughly analyse the properties of this model, {\it per se}. Since it
represents bubbles as soft disks (in the 2D case) it has some relevance to granular materials as
well.
\begin{figure}[htbp]
\begin{center}
\includegraphics[height=1.8cm]{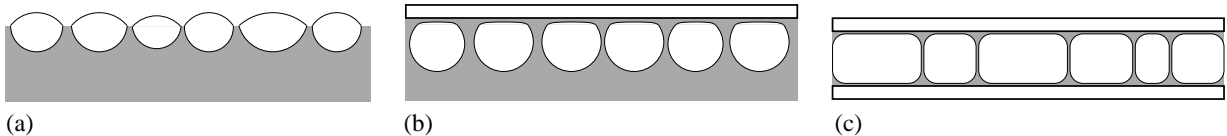}
\caption{Three types of 2D foams: 
{\it (a)} monolayer of air bubbles sitting at an
air/liquid interface (Bragg raft);
{\it (b)} bubbles floating in liquid under a glass plate
{\it (c)} bubbles confined
between two glass plates.
There are large effects due to the drag associated with motion relative to
solid boundaries in both (b) and (c). 
}
\label{2dfoams}
\end{center}
\end{figure}
As is often the case in foam physics, this study will remain for the time being in two dimensions,
which has obvious advantages. The  main experimental literature of foam rheology is concerned with
ordinary three-dimensional (3D) foams, but in recent years considerable
attention has been focused on their 2D
counterparts.

It has turned out that the obvious 2D experimental sample, consisting  of foam
trapped between two plates (see Fig.~\ref{2dfoams} (c)), has shear properties
that are significantly affected by viscous drag forces
exerted by the confining plates. We will therefore be concerned with two quite
different
but related cases:
with and without such forces. The latter case can be realised
experimentally in 2D as a Bragg raft (figure \ref{2dfoams}(a))
and it is roughly analogous to a 3D foam, because of the absence of
confinement-induced forces. Having defined the model, we will deal
with this case
first, and proceed to introduce the wall forces at a later stage. In summary the
main goals of this work are:
\begin{itemize}
\item to extract the parameters of a continuum (Herschel-Bulkley)
formalism, by means of simulation;
\item to use these in a continuum model for 2D shear localisation.
\end{itemize}
Both of these goals are satisfactorily realised withing the scope of the
present calculations.

\subsection{Questions raised by experiments}
It is necessary first to review some of the history of foam rheology.
The more traditional 3D experiments, for example those of Khan {\em et al.} \cite{
KhanEtal88}, have used various
types of rheometers to explore  the relation between stress, 
strain and strain rate. This is most
straightforward when strain rate does not change sign, that is, increments of shear are always in
the same sense, and we shall adopt that restriction here also. Awkward complications arise from
hysteretic effects in the more general case \cite{MarmottantGraner07}. The
conclusion has generally been that the data is well accounted for by the
Herschel-Bulkley equation. This adds to the quasistatic stress-strain relation a
second term proportional to the strain rate $\dot{\gamma}$, raised to the power
$a$,
\begin{equation}
 \sigma - \sigma_y = c_v \dot{\gamma}^{a},
\label{e:Herschel-Bulkley}
\end{equation}
where $\sigma$ and  $\sigma_y$ denote respectively stress and yield stress,
$c_v$ is the viscosity component of stress 
(also called {\em consistency}), and $a$ is the
Herschel-Bulkley exponent. However, the exact value of $a$ is still subject to
debate. Most experiments
agree on the shear-thinning behaviour of foams: whereas for $a=1$
(the Bingham fluid) the effective viscosity (that is, stress divided by strain
rate) tends to a {\em constant} at high rate of strain, it tends to {\em zero}
if $a<1$. While Schwartz and Princen \cite{Schwartz87} predicted
theoretically that $a=2/3$, various values between $0.25$ and $1$ have
been measured experimentally (see for instance
\cite{Lemlich85,KhanEtal88,Princen89,Gopal99,Lauridsen04,Denkov05,
KatgertEtal07}).
Besides, Denkov
\textit{et al.} \cite{Denkov05} and Katgert \textit{et al.}
\cite{KatgertEtal07} showed respectively
that $a$ depends on the properties of the surfactant and on 
the polydispersity of the foam. The obvious question  arises: what determines
the value of $a$? There has been little understanding of this up to now, and it
is one of our
stated objectives to explore the question within the context of a simple
model.

A second question concerns normal stresses or the related phenomenon of dilatancy. While often
 mentioned in the context of granular materials, these effects have not been much considered for
 foams, apart from the work of Weaire and Hutzler \cite{Weaire03,Rioual05},
confined to the quasi-static case. As they noted,
 there is in principle a second and possibly more important {\em dynamic} 
contribution to these
 effects,
 well known to rheologists \cite{Bagnold41,Wolf96}. So one may ask: what are the dynamic effects, in a
 simple model?

As in much of foam physics, recourse may be made to two-dimensional
systems, for the sake of simplicity and transparency. We will
not attempt a full review of the various rheological experiments on 2D foams
that have been performed over the last five years
\cite{Debregeas01,Lauridsen04,Wang06,KatgertEtal07}. Equally, we
will give no details of the quasistatic calculations
\cite{Kabla03,WynEtal08} that have been
adduced to account for them, or the continuum theory
\cite{Janiaud06,ClancyEtal06,JaniaudEtal07,WeaireEtal08b} which has
offered an alternative viewpoint. We will argue that the present
paper strongly supports the continuum model. Some results of the
quasistatic calculations suggest that corrections are needed to
account for the role of polydispersity, ad we do not rule that out.

\subsection{The soft sphere/disk model}
\label{s:model}

\begin{figure}[htbp]
\begin{center}
\includegraphics[height=5cm]{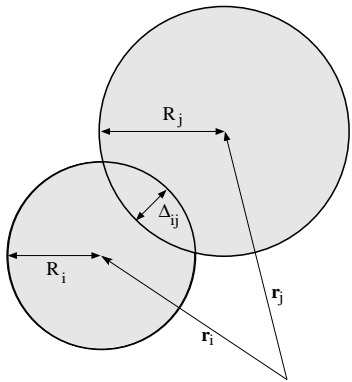}
\caption{Overlap $\Delta_{ij}$ between two contacting bubbles of radii
$R_i$ and $R_j$, located at ${\bf r_i}$ and ${\bf r_j}$, respectively.
}
\label{collision}
\end{center}
\end{figure}

In the model developed by Durian around 1995, 2D bubbles are represented by circular disks. When
overlapping (and only then) they interact via a simple spring force, 
the displacement of the spring
being the radial overlap (see Fig.~\ref{collision}). The elastic repulsive force
$ {\bf F_n}$ acting on bubble $i$ due to bubble $j$ is given by
\begin{equation}
\label{eq:fn}
 {\bf F_n} = \kappa \frac{2 R_0}{R_i+R_j}\Delta_{ij}
{\bf n_{ij}}.
\end{equation}
Here $\kappa$ is the coefficient of elasticity,
${\bf n_{ij}}$ is the normal vector between bubbles $i$ and $j$,
\begin{equation}
{\bf n_{ij}} = \frac{{\bf r_i}-{\bf r_j}}{ \vert{\bf r_i}-{\bf
r_j}\vert},
\end{equation}
and the overlap $\Delta_{ij}$ (see Fig.\ref{collision}) is given by 
\begin{equation}
\Delta_{ij} = \left\{ \begin{array}{rl}
(R_i+R_j) - \vert{\bf r_i}-{\bf r_j}\vert  &\mbox{if }(R_i+R_j) < \vert{\bf
r_i}-{\bf r_j}\vert \\
 0 &\mbox{otherwise.}
\end{array}
\right.
\end{equation} 
$R_i$ and $R_j$ are the 
radii of bubbles $i$ and $j$, centred at ${\bf r_i}$ and ${\bf r_j}$, respectively, and 
$R_0$ is the average bubble radius of the entire bubble packing. 
The ratio $\frac{2 R_0}{R_i+R_j}$ in eqn.~\ref{eq:fn} takes into account that
larger bubbles
are easier to deform than smaller ones. In such a model one may define an
effective liquid fraction
$\phi$ (which ignores the overlap of disks) as 
$\phi ~=~ 1 ~-~ N \langle {R_0}^2 \pi\rangle/A$, where  $A$ is the
area of the confinement of the disks and $N$ is the total number of disks. 
Since a packing of polydisperse disks loses its mechanical rigidity for $\phi
> 0.16$, for higher values of $\phi$ it no longer represents a
two-dimensional foam. In all the following, the liquid fraction will be chosen
as $\phi = 0.05$.

A real flowing foam dissipates energy by viscous friction in the
films and Plateau borders separating the bubbles. 
The films are not explicitly represented in our
model. The simplest expression, as used by Durian \cite{Durian97} and
adopted here, 
represents the
viscous  force ${\bf F_d}$ on bubble $i$ associated  with a neighbouring bubble
$j$ as
\begin{equation}
 {\bf F_d} = - c_b ({\bf v_i}-{\bf v_j}) 
\end{equation}
where $c_b$ is the dissipation constant for the bubble-bubble interaction
and
 ${\bf v_i}$ and ${\bf v_j}$ are the respective bubble velocities. With all the
above definitions, the model can provide a semi-quantitative description 
of foams throughout the range of liquid fraction consistent with stability. 
%
%
\begin{figure}[htbp]
\begin{center}
\includegraphics[height=5cm]{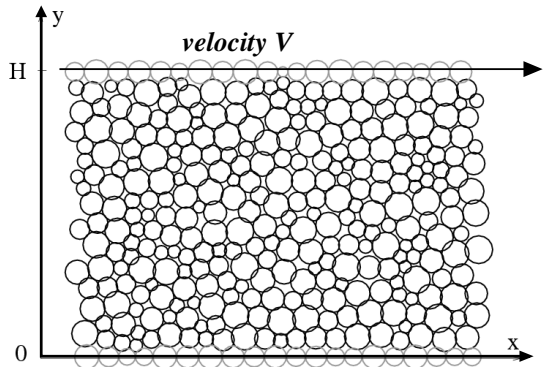}
\caption{Snapshot of 2D soft disk foam. The sample is  
periodic in the horizontal direction, bubbles at the upper and lower
boundary are shown in gray. In the simulation the upper boundary is moved
with velocity $V$.}
\label{snapshot}
\end{center}
\end{figure}
In Durian's original calculations \cite{Durian97}, inertia was neglected.
Hence all forces on each bubble were balanced. This reduces the problem to a set
of linear equations in the velocities of the bubbles. Durian then further 
simplified the problem by substituting the viscous drag exerted on each bubble
by its neighbours to the value the drag would have in a linear velocity profile.
He thus sets the average velocity $\langle{\bf v}_j\rangle$ of all neighbours
$j$ of bubble $i$ to $\langle{\bf v}_j\rangle = \dot{\gamma} y_i \hat{\bf x}$
where $\dot{\gamma}$ is the imposed strain rate and $y_i$ is the y-coordinate of
bubble $i$. 

In our calculations we instead allow each bubble to move independently,
subject to the elastic and dissipative forces defined above. In practice we
used the Verlet algorithm, with a bubble mass $m_b$ small enough to 
assume that inertial effects are negligible (the ratio $\kappa m_b/c_b^2$ was
set to 0.01). We present results here only for the eventual steady state after
long times. The results obtained are significantly different from those of
Durian. This
may be due to the strong approximation that he used, as indicated above. This
approximation had been removed later by Tewari \textit{et al.} \cite{Tewari99},
but they were not interested in the rheological properties and concluded that
both versions of the model were equivalent.

\section{Response to simple shear in the absence of wall drag}
\subsection{The Herschel-Bulkley power law exponent}
\label{s:rheology}
In this section we deal with the Durian model, with no additional viscous
drag force from confining plates.
Our first computations concerned the evaluation
of its
flow properties under simple shear, in particular the value of the
Herschel-Bulkley power-law
exponent $a$ in eqn.~(\ref{e:Herschel-Bulkley}).
To this purpose we generated assemblies of $1,000$
to $10,000$ bubbles in a rectangular confinement, as shown in
figure~\ref{snapshot}, using periodic boundary conditions in the horizontal
direction. The bubbles at the upper boundary, which are treated as
attached to this boundary, are given a constant
velocity $V$. 
This corresponds to the application of strain at a constant rate
$\dot{\gamma}
= V/H$ for a sample of width $H$, see Fig.
\ref{snapshot}. We will only consider polydisperse foams, 
the bubble radii having a uniform
distribution within the range $R=R_0(1\pm0.15)$.

At this stage it is convenient to introduce a dimensionless Deborah 
number $De$ which is defined as the ratio of the characteristic time
of the material that is sheared to the characteristic time of the
deformation process \cite{BarnesEtal89}. We identify the latter with
$\dot{\gamma}^{-1}$, and the material time-scale we
relate to
the competition of energy storage and dissipation at the level of
bubble-bubble interactions. This then results in the definition
$\textrm{De} = \dot{\gamma}
c_b/\kappa$.

\begin{figure}[htbp]
\begin{center}
(a)\includegraphics[height=6cm]{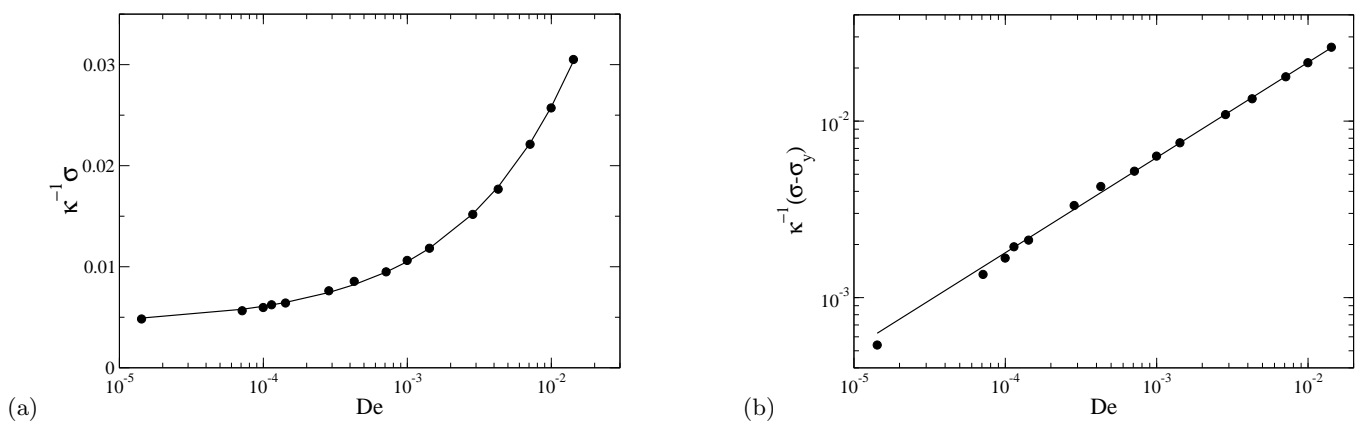}
\hfill
(b)\includegraphics[height=6cm]{fig4b.ps}
\caption{(a) Stress on the moving boundary as a
function of the Deborah number $\textrm{De} = \dot{\gamma}c_b/\kappa$. The
solid line is a fit to eqn. \ref{e:Herschel-Bulkley} (re-expressed in
terms of $\textrm{De}$),
resulting in the Herschel-Bulkley exponent $a= 0.54 \pm 0.01$.
In (b) we have subtracted the fitted value of the yield stress $\sigma_{y}$ from the data to show the power-law behaviour in a double
logarithmic plot.}
\label{rheology}
\end{center}
\end{figure}

Under an imposed boundary shear,
after a transient, the system of disks reaches a
steady average state. It is characterized by a linear average velocity profile 
of the disks
with regard to their vertical position in the sample, so there is no
localisation in the present case; we may proceed to extract the
constitutive law in a straightforward way. We determine the 
stress at the moving boundary
as a function of $V$. In 
figure~\ref{rheology} we display this variation as a function of the
dimensionless Deborah number $\textrm{De}$.
A least square fit of the data with the Herschel-Bulkley form results in 
\begin{equation}
\sigma/\kappa = 0.0043 + 0.26 \,\text{De}^{(0.54\pm 0.01)}.
\label{e:hb}
\end{equation}
The model foam thus exhibits a strongly non-linear, shear-thinning rheological
behaviour, despite the linearity of all local forces. This can be attributed to
the importance of disorder in such a polydisperse jammed system: the velocity
fluctuations cannot be neglected and make the trajectories of the
bubbles strongly non-affine. Therefore the simple image
of bubble layers sliding over each other is misleading. In
the initial version of the model applied by Durian \cite{Durian95,Durian97}, the
mean-field approximation effectively substracts these disordered motions,
which results in a Bingham rheology ($a=1$). Let us note that the non-linear
behaviour
does not affect the {\em average} velocity profile in the simple shear geometry
we adopted (as opposed to what happens in a cylindrical Couette geometry
\cite{Lauridsen04}). 
The shear localisation that is discussed in later
sections in not seen here.

The value of the Herschel-Bulkley exponent we obtain
(eqn.~(\ref{e:Herschel-Bulkley})) 
$a= 0.54 \pm 0.01$ is roughly consistent with the various
experimental measurements already mentioned, that showed a nonlinear,
shear-thinning
behaviour \cite{Lemlich85,KhanEtal88,Princen89,Gopal99,Lauridsen04,Denkov05,
KatgertEtal07}. However, it
is slighly higher than most experimental values. This
discrepancy can be attributed to the extreme simplicity of the local
ingredients of the model that we have used here. Including more realistic forces
at the local scale might lead to a better estimation of the Herschel-Bulkley
index. Recent theoretical work by Denkov {\it et al.} \cite{Denkov08} implies
that the viscous dissipation between two bubbles sliding along each other
should scale like $\Delta V^{1/2}$ rather than being linear. However, the
relation between the local properties and the macroscopic rheology is still not
understood. Further work on this topic is currently performed and will be the
subject of a subsequent publication.

\subsection{Normal stress}
\begin{figure}[htbp]
\begin{center}
(a)\includegraphics[height=6cm]{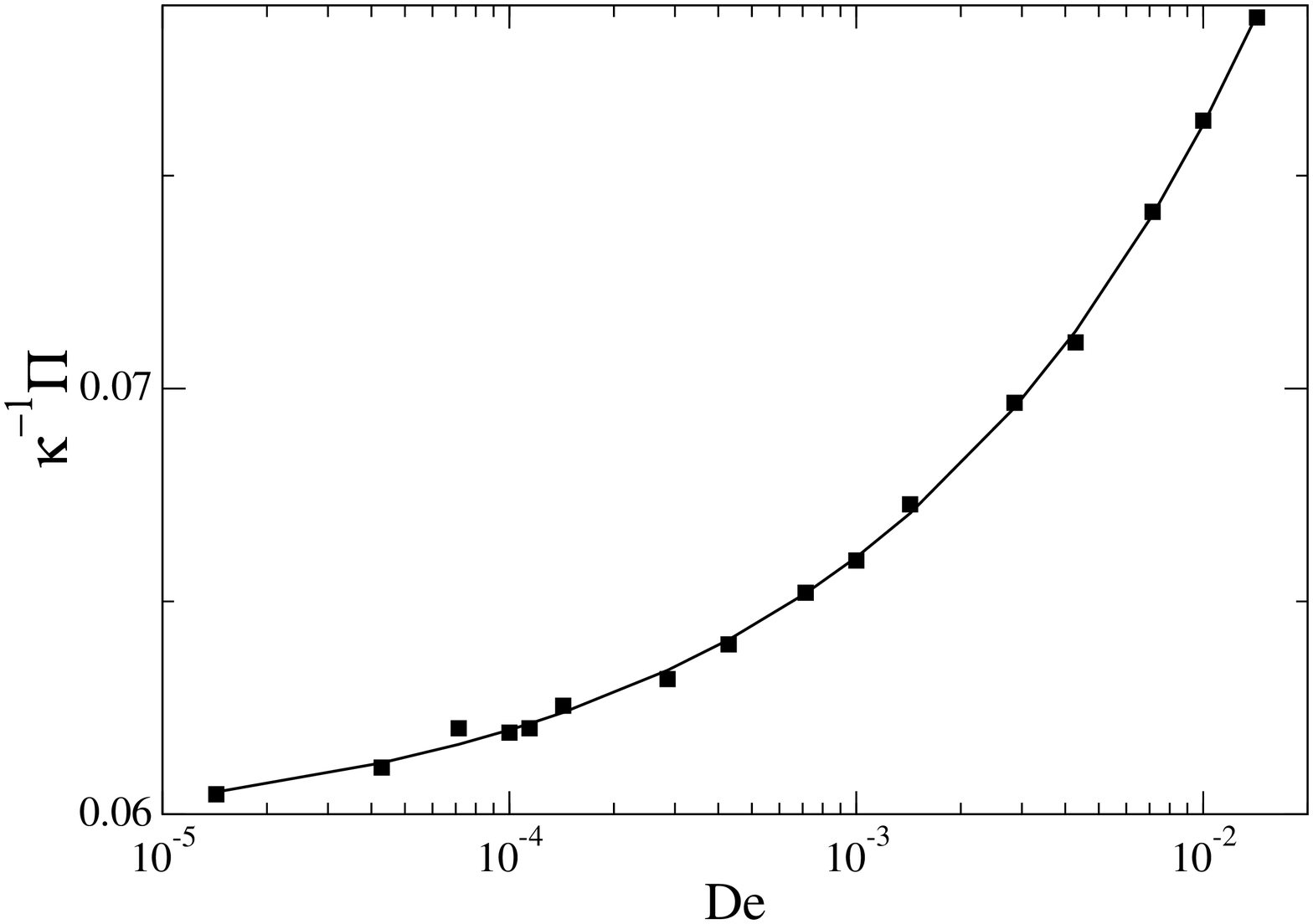}
\hfill
(b)\includegraphics[height=6cm]{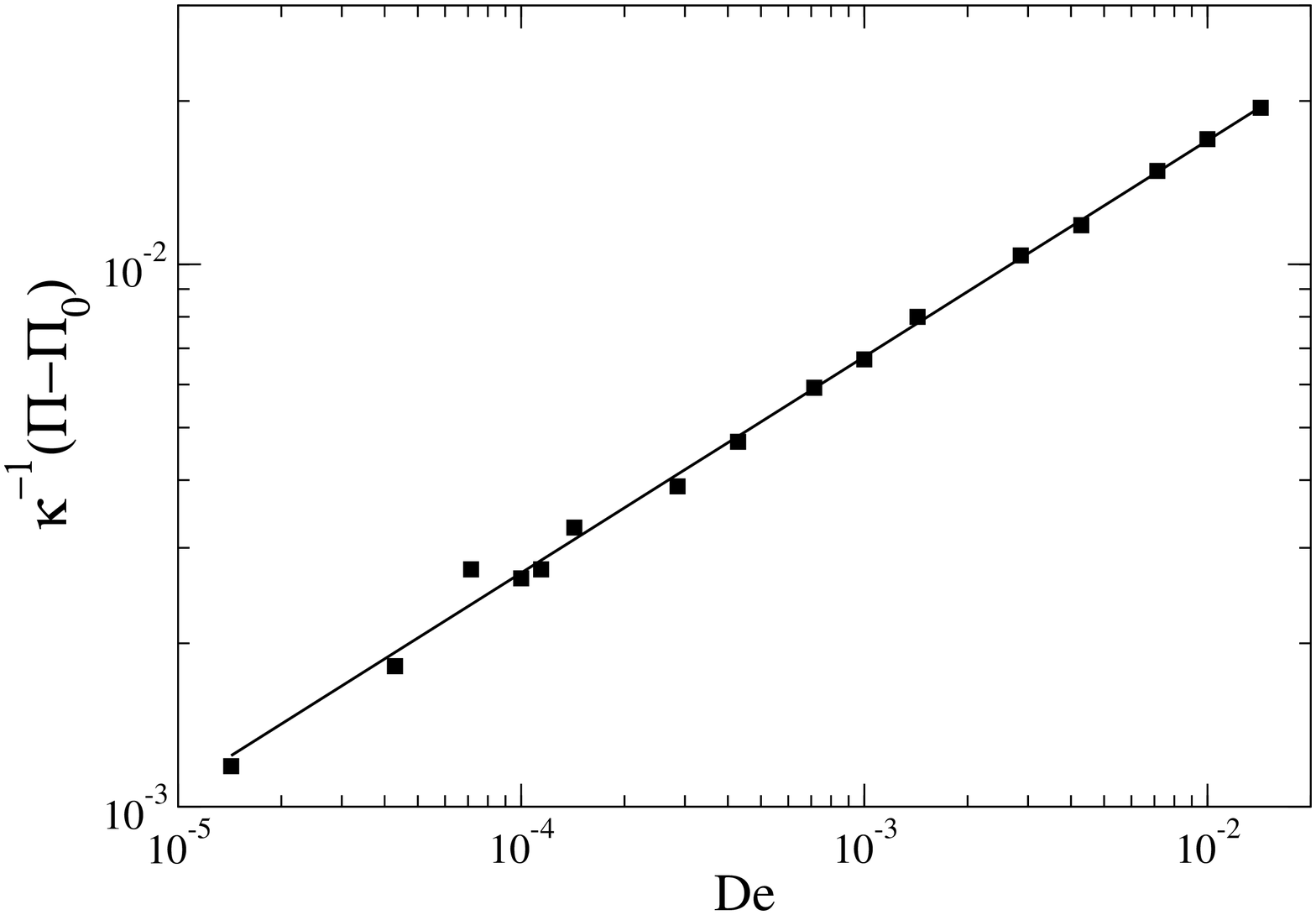}
\caption{Pressure acting on the moving boundary as a
function of the Deborah number. The data is well described by
eqn.(\ref{e:osmotic}) with an exponent $0.4 \pm 0.01$. In (b) we have
subtracted the fit parameter $\Pi_0$ to show the power law behaviour in a
double-logarithmic plot.}
\label{dilatancy}
\end{center}
\end{figure}

In a system constrained by a fixed applied pressure, the foam would expand 
in volume (by
increasing its nominal liquid fraction) when sheared. 
In our simulations we impose the volume by
fixing the position of the edge bubbles, and measure the resulting pressure
$\Pi$ on the boundaries. Its
variation with shear rate is shown in figure~\ref{dilatancy}
and is well described by a formula analogous to the Herschel-Bulkley
equation,
\begin{equation}
\Pi/\kappa = 0.059 + 0.11 \textrm{ De}^{(0.40\pm 0.01)}
\label{e:osmotic}
\end{equation}
For the case of $\textrm{De} = 0$ one obtains the 
static osmotic
pressure of the foam $\Pi_0$. 
Although the tendency of a 3D foam to
increase its liquid fraction under shear has been qualitatively reported in
experiments \cite{MarzeEtal05}, we do not know of any quantitative experimental
measurements of the dynamic variation of the osmotic pressure, which is
particularly difficult to obtain in 2D. Rheologists sometimes cite the old work
of Bagnold \cite{Bagnold41} as indicating a {\em
quadratic} dependence, very different from the power law in equation
(\ref{e:osmotic}). We have not proceeded any further with the analysis of the
normal stress.

\section{The effect of wall drag}
\subsection{Adding wall drag}

As we have already mentioned, certain experiments with a 2D foam between
two glass plates exhibit a new type of flow; when one boundary is moved to 
produce shear, the resulting shear is exponentially localised at that
boundary \cite{Wang06}. According to the continuum model of Janiaud {\it et al.}
\cite{Janiaud06,ClancyEtal06,JaniaudEtal07} this is to be understood as a direct
effect of the wall drag which we will now introduce into the numerical
model. This continuum model in its original form assumed a constitutive equation
of the Bingham type ($a=1$), and added in the drag force $F_w$, as a body force
proportional to local velocity. This predicted an exponential localisation of
the
flow along the moving boundary, the width of the shearband being independent of
the driving velocity.

To mimic that theoretical model in our simulations, we now add to 
the forces acting on bubble $i$ a wall drag force ${\bf F_w}$, given in the
most simple form by
\begin{equation}
 {\bf F_w} = -  c_w \vert {\bf v_i} \vert^b \frac{{\bf v_i}}{\vert {\bf v_i}
\vert} \label{e:wall_drag}
\end{equation}
where $c_w$ is a drag constant. According to Bretherton \cite{Bretherton61}
and Denkov {\it et al.} \cite{Denkov05}, for surfactants with low surface
viscosities, the friction of the foam due to the wall is characterized by
$b=2/3$.
However, to keep all the ingredients of our model linear, we will adopt $b=1$. 
The resulting
equation of motion is again
solved numerically, using the second order Verlet method.
Note that we have just established in section \ref{s:rheology} that the 
appropriate constitutive law is not that of Bingham, so we will have cause to
return to this point again. As we shall see, the
results for the steady shear at long times exhibit localisation, which
conforms well to the prediction of
the continuum model.

\subsection{Flow localisation}
\label{ss:wall_drag}
\begin{figure}[htbp]
\begin{center}
\includegraphics[height=6cm]{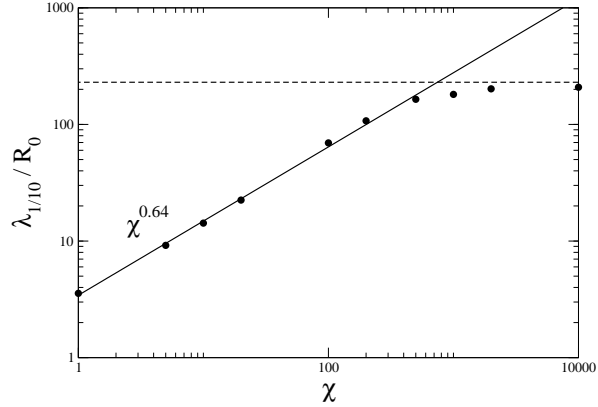}
\caption{Localisation length $\lambda_{1/10}$ as a function of the dimensionless ratio $\chi =
c_b/c_w$. The solid line is a fit of
the data for $\chi < 500$ to a
power law, resulting in 
$\lambda_{1/10}/R_0 \propto \chi^{0.64\pm0.01}$. The power law behaviour is
limited by the
width of the sample which is indicates by the dashed line (in units of the
average bubble radius
$R_0$).
}
\label{shearlength}
\end{center}
\end{figure}

Shear simulations with added wall drag
were performed on an  
assembly of $10,000$ soft disks, for constant values of $\chi=c_b/c_w$, the
ratio of dissipation and drag constant. Once
a steady state was reached, we determined the velocity profile between
static and moving boundary by performing time averages over the horizontal
velocity components of the bubbles.
\begin{figure}[htbp]
\begin{center}
\includegraphics[height=8cm]{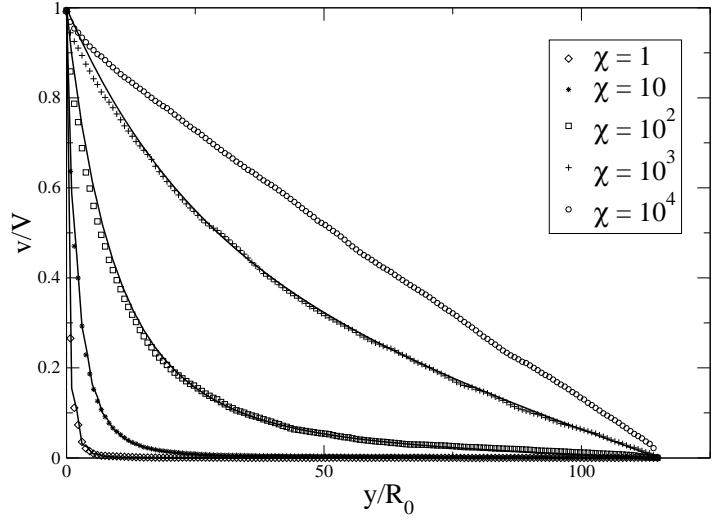}
\caption{Normalized velocity profiles in the foam for a range of 
values of the parameter $\chi=c_b/c_w$. 
Shear close to the moving boundary is enhanced as wall friction is increased,
i.e. with decreasing $\chi$. The solid lines, which agree closely with the data,
were obtained numerically from the continuum model, using the Herschel-Bulkley
index $a=0.54$.}
\label{shearbands}
\end{center}
\end{figure}
While
in the absence of wall friction ($c_w = 0$, i.e. $\chi \rightarrow \infty$), the velocity profile is 
roughly linear throughout the sample, this is no longer the case for finite
values of $\chi$. Instead flow is 
localised near the moving boundary, as shown
in figure \ref{shearbands}.
A localisation length  $\lambda_{1/10}$  can be arbitrarily defined 
by measuring the distance from the
moving boundary at which the
velocity reaches 1/10th of its value at the boundary (see also Appendix
\ref{a:def_length}). The variation of $\lambda_{1/10}$
with $\chi$ is shown in 
figure~\ref{shearlength}. For values of $\chi \lesssim 500$ (i.e., if the
shearband width is less than the sample width $H$) the localisation length
varies with
$\chi$ in the form of a power-law, $\lambda_{1/10} \propto \chi^{0.64\pm 0.01}$.

\begin{figure}[htbp]
\begin{center}
\includegraphics[height=6cm]{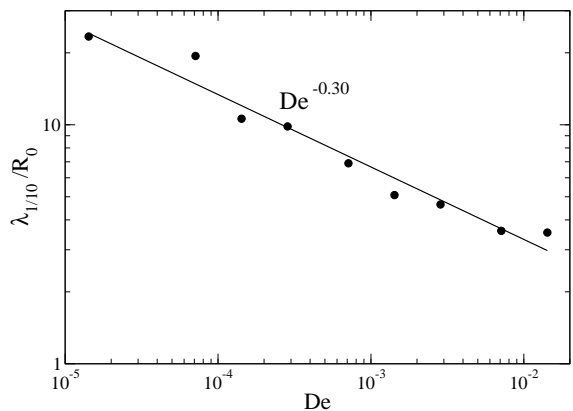}
\caption{Normalised localisation length $\lambda_{1/10}/R_0$ as a function of the
Deborah number. The solid line is a fit of
the data to a power law, resulting in 
$\lambda_{\frac{1}{10}}/R_0 = 0.87\, \mbox{De}^{-0.30}$,
in excellent agreement with theory. (Here the
sample width was 28 $R_0$).}
\label{shearlength_vel}
\end{center}
\end{figure}

Figure \ref{shearlength_vel} shows the evolution of $\lambda_{1/10}$ with the
Deborah number De, when varying the driving velocity $V$. A least square fit
results in 
\begin{equation}
\lambda_{1/10}/R_0 = 0.87\, \mbox{De}^{-0.30}.
\label{e:scaling_sims}
\end{equation}

Such a power-law scaling can also be found theoretically by modifying the
continuum model of Janiaud \textit{et al.} to incorporate the
Herschel-Bulkley rheology we exhibited here: the velocity field $v(y)$ 
has to be such that the internal dissipation is exactly balanced by the friction
along the confining plates, that is
\begin{equation}
 c_v \frac{d}{dy}\left[\left(\frac{dv}{dy}\right)^a \right] = c_d v(y)^b \, .
\end{equation}
Here $c_d$ is the wall drag constant $c_w$ of eqn.~\eqref{e:wall_drag} per
unit area, i.e. $c_d = c_w/(\pi {R_0}^2)$ and $b=1$ for the case of the simple
linear form for the drag force considered here. The exponent $a$ is the one
from the Herschel-Bulkley relationship, i.e. $a=0.54$ in our case (see eqn.
\ref{e:hb}), and $c_v$ is the viscosity component of stress. This equation can
 be solved to predict the localisation of the flow, as shown theoretically in
\cite{WeaireEtal08b}:
the (exponentially defined) localisation length $l$ is given by
\begin{equation}
l \simeq \left(\frac{c_v}{c_d}\right)^{\frac{1}{1+a}} V ^{\frac{a-b}{1+a}},
\label{e:loc_length_cont}
\end{equation}
provided that $l$ is much less than the sample size (the two definitions of the
localisation length are related by
$\lambda_{1/10} = \ln (10) \; l$, see Appendix \ref{a:def_length}). This
immediately gives the scaling $\lambda_{1/10} \propto {c_w}^{0.65}$, in
excellent
agreement with the data shown in figure \ref{shearlength}. However, the
success of the continuum model is not confined to such scaling relationships,
but is fully quantitative. Rewriting eqn.~(\ref{e:loc_length_cont}) in terms of
the Deborah number gives
\begin{equation}
 l \simeq \left(\frac{c_v}{c_d}\right)^{\frac{1}{1+a}}
\left(\frac{H \kappa}{c_b}\right)^{\frac{a-b}{1+a}}
\textrm{De}^{\frac{a-b}{1+a}}.
\end{equation}
Inserting
our simulation input parameters for $c_d, H, c_b, R_0$ and $b=1$, together with
our numerical results for the values of $c_v$ and $a=0.54$, we obtain 
\begin{equation}
l/R_0 = 0.36 \mbox{De}^{-0.30}. 
\end{equation}
This corresponds to $\lambda_{1/10}/R_0 =
0.83 \mbox{De}^{-0.30}$, which is also in excellent agreement with the numerical
results.
We have concentrated on the localisation length here, but the full velocity
profile, which turns out to be only approximately exponential, can be
numerically calculated from it. As figure \ref{shearbands} shows, the result is
in close agreement with the data.

\section{Conclusions}
\label{s:conclusions}
By means of numerical simulations based on the soft-disk model, we have shown
that the Herschel-Bulkley rheology of a 2D foam can be derived from a discrete
model with good agreement with experiments, which links for the first time the
continuum and local descriptions of a foam. Our model also predicts a dynamic
dilatancy, that is, a tendency of the foam to increase its volume when sheared.
Finally, we added to the classical bubble model the viscous friction
experienced by a foam under confinement, which resulted in the formation
of shearbands, as had been observed in experiments and derived with a
continuum model. Adapting this rheological model to our parameters lead to a
consistent picture of the link between the local and global descriptions. This
must be regarded as a strong indication for
the theory, but it will be important to examine the limits of that
conclusion.


Previous quasistatic simulations \cite{Wyn08} suggest that disorder
(polydispersity) plays a role. So far, our interpretation 
is that wall drag responsible for localisation, as prescribed by the continuum
model, but that the eventual localisation length contains an {\em
additional} contribution from polydispersity, not contained within the
continuum description.

In further studies, we will explore this by concentrating on the limit in
which $\lambda \rightarrow 0$ for the continuum model, and seek to identify
the effect of polydispersity, which is expected to be significant in that
limit. It will also be informative to repeat simulations with different local
forces, to try to establish the precise relation (if any) of the
Herschel-Bulkley parameters to these forces. Finally, if our conclusions are
sustained for this and similar models, the core theoretical problem of the
origin of the Herschel-Bulkley nonlinearity will remain.

\appendix
\section{Definition of localisation length}
\label{a:def_length}
The precise definition of localisation length is in general arbitrary. 
For the analysis of our numerical results we chose to use
\begin{equation}
v(\lambda_{1/10}) = V/10,
\end{equation}
where $v(x)$ is the local average bubble velocity, measured a distance $x$
away from the boundary which moves at velocity $V$.

In the continuum theory an alternative definition was used,
\begin{equation}
v(l) = V/e,
\end{equation}

There is no general relation between these two, but since localisation is in
general {\em approximately} exponential,  we may use
\begin{equation}
\lambda_{1/10} \simeq \ln10 \; l \simeq 2.30 \; l.
\end{equation}
to adjust the theoretical prediction. We have adopted this procedure in
section \ref{ss:wall_drag}.

\acknowledgments
This work was supported by the European Space Agency (MAP
AO-99-108:C14914/02/NL/SH and (MAP
AO-99-075:C14308/00/NL/SH) and Science Foundation
Ireland (RFP05/RFP/PHY0016). We also acknowledge the European Union COST P21 action on "The
Physics of droplets". V.J. Langlois  would like to acknowledge the Grenoble Foam Mechanics
Workshop and to thank G. Katgert and M. van Hecke for fruitful discussions, and for providing some
of their results before publication.


\end{document}